%% file: main.tex
\RequirePackage[hyphens]{url}

\documentclass[runningheads]{llncs}

\usepackage[T1]{fontenc}

\usepackage{graphicx}
\usepackage[english]{babel}

\usepackage{float}
\usepackage[affil-it]{authblk}
\usepackage{xcolor}

\begin{document}

\date{}

\title{Wallcamera: Reinventing the Wheel?}

\author{Aur\'elien Bourquard\inst{1} \and
Jeff Yan\inst{2} 
}

\authorrunning{Bourquard \& Yan}

\institute{Massachusetts Institute of Technology, USA\\
\email{aurelien@mit.edu} 
\and
University of Southampton, UK \\
\email{jeff.yan@soton.ac.uk}
}

\maketitle

\begin{abstract}
The Wallcamera \cite{iccv21}, developed at MIT CSAIL, has captivated the public's imagination.  
Here, we show that the key insight underlying the Wallcamera
is the same one that underpins the concept and the prototype of \textit{differential imaging forensics} (DIF), both of which were validated and reported several years prior to the Wallcamera's debut.

Rather than being the first to extract and amplify invisible signals---aka \emph{latent evidence} in the forensics context---from wall reflections in a video, or the first to propose activity recognition \emph{per se} following that approach, the Wallcamera's actual innovation is achieving activity recognition at a finer granularity than DIF demonstrated. 

In addition to activity recognition, DIF as conceived has a number of other applications in forensics, including 1) the recovery of a photographer's personal identifiable information 
such as body width, height, and even the color of their clothing, from a single photo, and 2) the 
detection of image tampering and deepfake videos. 

\end{abstract}

\input{sections/introduction}

\input{sections/results}

\input{sections/comparisons} 

\input{sections/conclusion}

\bibliographystyle{unsrt}

\bibliography{bibliography}

\end{document}

%% file: sections/introduction.tex
\section{Introduction}

It recently came to our attention that 
an interesting idea called the Wallcamera (Sharma et al \cite{iccv21}) was proposed by a distinguished team   
at the 
Computer Science and Artificial Intelligence Laboratory (CSAIL), MIT.
Entitled `\textit{What you can learn by staring
at a blank wall}' and published at a leading international venue for computer vision, 
their work
also captured the imagination of the public. For example, it was featured by the Scientific American 
\cite{SciAm}.

More than two years before the Wallcamera, we published a paper \cite{dif} entitled `\textit{Differential Imaging Forensics}' (DIF) 
at \url{arxiv.org} across different technical fields including Cryptography and Security (cs.CR); Computer Vision and Pattern Recognition (cs.CV); and Multimedia (cs.MM).

The paper titles might suggest that these were
entirely different research. However, we were stunned by some major similarities between them. 
In this 
article, we address a simple question: Is the Wallcamera a reinvention of DIF?
Our answer is both yes and no. On the one hand, the key ideas behind the two papers were exactly the same, and the experimental settings were virtually identical. On the other hand, it is fair to recognise that the Wallcamera achieved a finer granularity of activity recognition than DIF did. However, unlike the Wallcamera, DIF as conceived entails a lot more than activity recognition.

%% file: sections/results.tex
\section{Concepts and Results}
\label{Sect: Results}

\subsection{DIF}

The term \emph{differential imaging forensics} (DIF) was first coined by Bourquard and Yan \cite{dif} in 2019. 
Motivated by the simple question: \emph{`Given a single photo, 
how can we determine who was behind the camera?'}, this new area of research had emerged even earlier, with preliminary but promising results \cite{yan2017poster} published as early as 2017. 

DIF entails groundbreaking forensic techniques to uncover evidence that is readily available in an image or video footage but would otherwise remain faint or invisible to a human observer. Specifically, it allows for the computational extraction and amplification of visual evidence---such as dim reflections---created by subtle interactions of light occurring inside a scene. This is achieved by conducting a comparative analysis between an image of interest and an additional \emph{reference baseline image}, the latter being acquired under similar conditions, but ensuring that potential individuals or objects whose evidence is to be uncovered are absent from the scene. Bourquard and Yan \cite{dif} successfully demonstrated the effectiveness of this differential-imaging paradigm through successful experiments with both images and video footage.

The DIF methodology thus involves two stages. The first stage is the acquisition of the reference baseline image, which complements the image of interest from which visual evidence is to be retrieved. This acquisition stage is most straightforward in video settings where all scene frames are acquired from the same camera position and viewpoint. The second stage is differential image analysis, where the differences between the image (or video frame) of interest and the corresponding reference baseline image (or frame) are extracted and amplified computationally, thereby becoming perceptible to human observers.

\begin{figure}[H]
\centering
\includegraphics[width=\textwidth]{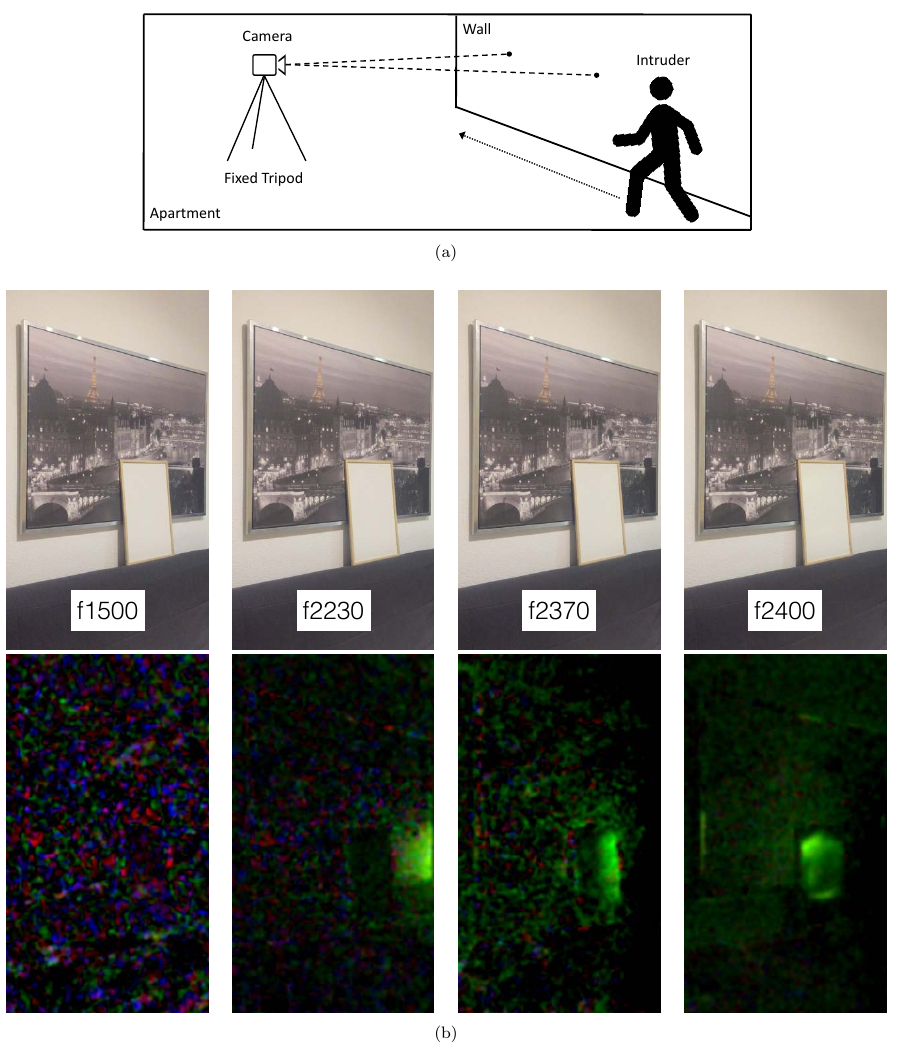}
\caption{The gist of DIF in a video-experiment setting similar to the Wallcamera's, reproduced here from Figure 3 in Bourquard and Yan \cite{dif} (published in 2019). (a) Schematics of the scene setting with a camera pointed towards the wall. The space between the wall and the camera is sufficient for the passage of an intruder in-between. (b) Available scene frames (top row) and corresponding difference scene frames (bottom row). From left to right: (1) the intruder stands outside the apartment with the door closed, (2) the intruder has just entered the apartment, (3) the intruder approaches a plastic white board, and (4) the intruder stands in front of the plastic-white-board surface. All the video frames and contents thereof are best seen magnified in an e-copy.}
\label{fig:ResultsDIF}
\end{figure}

Practical implementations of DIF are able to detect the presence or absence of a person in a room, who is located outside the field of view (FoV) of the camera. Figure \ref{fig:ResultsDIF} illustrates an experiment where DIF was applied to a video footage acquired by a camera looking at a wall \cite{dif}. This experiment demonstrated that DIF can provide specific information on an intruder, including biometrics such as body height and width, and non-biometrics like the color of their clothing. As far as the video camera and any bare human eyes were concerned, no details from the intruder were perceived, just like what the Scientific American article \cite{SciAm} vividly described years later. Moreover, this experiment clearly revealed when the intruder was absent from the room, when he was present, and when he was walking in the room. The experiment also provided information on the intruder's approximate locations throughout the relevant video frames.

\clearpage

\subsection{The Wallcamera}

Sharma et al. \cite{iccv21} demonstrated that by filming a blank wall in a room as people moved around inside, an observer outside the room, unable to see the people directly, could determine the number of people present and their activities. The proposed technique extracted weak signals created by the subtle impact of human bodies on the wall, which were imperceptible to the human eye, and then amplified these signals. Neural networks trained with these signals could infer whether zero, one, or two people were present and classify activities such as walking, jumping, waving hands, crouching, or no activity (i.e., all human subjects being static). This technique effectively turned a blank wall into a camera, enabling the deduction of the number of people and their activities in the room by observing the wall.

\begin{figure}[H]
\centering
\includegraphics[width=\textwidth]{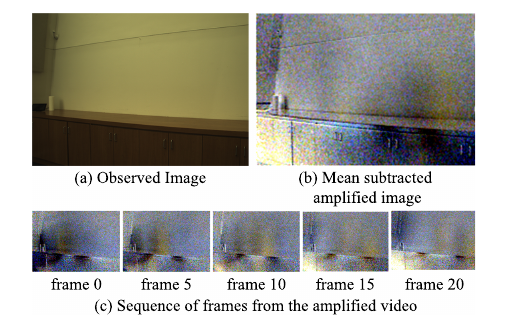}
\caption{The gist of the Wallcamera. Note: both this figure and the following caption are 
reproduced from Figure 3 in Sharma et al \cite{iccv21} (published in 2021).
(a) A representative frame of the seemingly static input video. (b) A frame of the amplified residual video after subtracting the mean frame reveals faint changes in illumination caused by motion of the people. (c) A sequence of frames shows the motion of these features. 
} 
\label{fig:ResultsWallcamera}
\end{figure} 

The Wallcamera experiment shown in Figure \ref{fig:ResultsWallcamera}, reproduced from 
Sharma et al \cite{iccv21}, illustrates the gist of their idea, explaining why it works. 
Figure 2(a) shows the wall as appeared to a human observer. It would appear static to the naked eye, due to the dominant contribution of ambient light to the light field in the room.  
Similar to the bottom row of Figure 1(b), Figures 2(b) and 2(c) demonstrate the subtle impact created on the wall by the light bouncing off those moving subjects in the room.

%% file: sections/comparisons.tex
\section{Comparisons and Discussions}

In light of the results shown in Section \ref{Sect: Results}, the DIF paper \cite{dif} was the first to demonstrate the extraction and amplification of invisible signals---aka \emph{evidence} in its forensics context---from wall reflections in a video. The practical experiment conducted in \cite{dif} also demonstrated the ability of DIF to perform some activity recognition. In this section, we discuss some key similarities and differences between DIF and the Wallcamera.

\subsection{Experiments: settings, configurations and cameras}

We first look into the similarities and differences between the experiments in DIF and the Wallcamera. 

Several key aspects of the experiments conducted therein were similar. Ambient lighting was assumed in both DIF and Wallcamera experiments. Furthermore, the objects or humans of interest were always located outside of the FoV of the camera, with no visible shadows thereof cast on the wall. 

The techniques proposed in both papers required no prior knowledge of the scene and no calibration. As described in Section \ref{Subsect: SignExtAmpl}, the methods used to extract and amplify the information of interest followed straightforward and well-defined steps, making them generic and scene-independent. Although the activity-recognition method developed for the Wallcamera was pre-trained on a fixed set of scenes, it can be used on new scenes without any retraining.

Some differences were as follows. In terms of where the latent but tale-telling signal (i.e., \emph{latent evidence} in the forensic context) was found, the Wallcamera considered walls only, at least in the experimental results reported thus far, whereas DIF was applied to different types of physical surfaces \cite{dif}. Moreover, the Wallcamera used a high-resolution camera (PointGrey
Grasshopper 3, at about US\$1,200 as of 2024), whereas DIF experiments only involved an iPhone 6 camera as default configuration. According to the Scientific American piece \cite{SciAm}, a standard digital camera created too much background noise for the Wallcamera, and the results based on smartphone cameras were even worse. 

\subsection{Non-line-of-sight (NLOS) aspect}

In conventional image-acquisition settings, objects of interest are located in the line of sight of the camera, so that the resulting image will contain them in its FoV. By contrast, non-light-of-sight (NLOS) settings involve objects that are not located in the direct line of sight of the camera. Instead, NLOS imaging exploits indirect optical information on the objects, exploiting light that reaches the camera via other paths through multiple scattering. Since the objects cannot be imaged directly, the increased flexibility of NLOS settings comes at the cost of an additional numerical-reconstruction procedure, which also requires the scattered light captured by the camera to contain enough relevant information on the hidden scene.

NLOS imaging techniques have given rise to a growing body of literature, and can be classified into active and passive methods. Active methods require to interact with and probe the environment, for instance using lasers to illuminate the hidden scene, thus allowing one to capture relevant information. By contrast, passive NLOS methods do not probe the environment, and only use camera sensors to capture the light information that is available in the default scene configuration. The NLOS concept can be applied to non-visible electromagnetic wavelengths and also extends beyond optical wavelengths, including other modalities such as WiFi or sound waves.

\begin{figure}[H]
\centering
\includegraphics[width=0.6\textwidth]{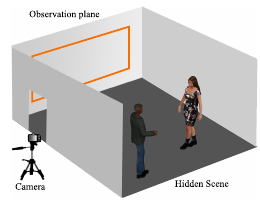}
\caption{Schematic setup of the Wallcamera, reproduced from Sharma et al \cite{iccv21} (i.e., its Figure 2a).
}
\label{fig:SetupWallcamera} 
\end{figure}

As shown in Figure \ref{fig:SetupWallcamera}, the Wallcamera experiment considered a passive NLOS setting, where the camera points towards an observation plane (i.e. the blank wall) that is located behind an open door, the scene of interest being hidden from the camera by a separating wall. 

In contrast, the video experiment in DIF, as depicted schematically in Figure 1(a), did not involve a deliberately concealed scene and might not immediately appear like a NLOS setting at first glance. However, during the entire DIF video experiment, the human subject walked outside the camera’s FoV and thus remained hidden from the camera’s perspective. This effectively makes the DIF video setting a NLOS scenario as well.

Importantly, the light interaction between the human subject and
the wall was similar for both the Wallcamera and DIF experiments; so
were the relative geometrical arrangements of the human subject, the wall, and the camera. It is worthwhile to emphasize that, while the aforementioned separating wall may affect the camera's FoV, its location never interfered with the light interaction between the blank wall and the human subject of interest, which thus remained similar to the DIF setting in any case. 

Furthermore, the Wallcamera work also involved experiments without the separating wall (see Section 6.4 in Sharma et al \cite{iccv21}), in which case their setting was the same as in our DIF experiment not only in essence, but also in actual overall appearance, except that they sometimes involved more than one human participants. 

We note that this section provides an intriguing connection between DIF and the NLOS research, and that this connection was not emphasised in Bourquard and Yan \cite{dif}. However, a major reason for conducting our DIF video experiment in this manner was simply to enable a controlled experiment that was both convenient and efficient. And all other application scenarios which we have conceived for DIF do not have to involve NLOS imaging.  

\subsection{Signal extraction and amplification}
\label{Subsect: SignExtAmpl}

The signal extraction and amplification method proposed in the Wallcamera 
\cite{iccv21}, while not exactly the same as the method used for DIF, shares several similarities with it. 

In both DIF \cite{dif} and the Wallcamera \cite{iccv21}, 
an initial extraction step is carried out to produce modified video frames whose content includes and isolates the specific differential visual information related to the persons or objects of interest to be recovered. In that regard, the Wallcamera  
\cite{iccv21} subtracts a unique frame, which is obtained as the temporal average of the whole video, from every raw frame of the video. When dealing with video data, the DIF paper involves a similar subtraction operation using a  
\emph{reference frame}, which is obtained as the temporal average of several---though not all---consecutive frames of the video \cite{dif}. The DIF paper also contemplated the possibility to use refined versions of that extraction step in further work, in which \textit{``temporal redundancy could be better exploited, and difference-image information could also be extracted from arbitrary (e.g., adjacent) frame pairs—as opposed to merely using the same reference image—to provide additional information"}. 

Following the extraction step described above, both DIF and the Wallcamera  
involve what amounts to a final signal-amplification step. This step maximizes the signal-to-noise ratio of the result, thus making it fit for visualization or further analysis. In both works, the key operations that provide signal amplification are signal averaging and contrast normalization. 

In essence, the amplified images resulting from the above operations for the Wallcamera are similar to the so-called \emph{difference images} in the DIF context \cite{dif}. It is these difference images that contain the tale-telling signals. In both cases, the key insight in extracting and amplifying invisible information from a video is to subtract one same reference frame or an averaged reference frame from every frame of interest to be analyzed, and then to use spatial filtering to amplify the signal, thereby producing the difference images. 

However, the specific implementation of these operations differ between both works. In the Wallcamera \cite{iccv21}, all images from the input video were first converted to the logarithmic space of values prior to any processing. This logarithmic-space conversion helped to equalize the magnitude of variations between light and dark regions of the scene. 
Next, temporally constant intensity components were  
removed by computing the temporal average of the video and subtracting it from each frame. To minimize noise and maximize contrast, averaging was then performed through spatial down-sampling in both horizontal and vertical dimensions. The resulting values were re-scaled to the full range of RGB values by multiplying them by a factor of 50 and adding a middle-gray base level to make the negative values visible. 

Next, an additional temporal correction, only used in the Wallcamera, was applied to address periodic intensity sways. These fluctuations were attributed by the authors to small discrepancies between the frame rate and the flickering of lights at the AC frequency (60HZ). The corrections were made as follows. First, the median light-intensity value was first computed for every video frame, creating a \emph{global} 1D curve that described the evolution of the average light intensity as a function of time. Then, at each pixel position of the video, the \emph{local} temporal intensity variation across frames was projected onto the orthogonal complement of this global curve, thus removing the effect of potential intensity sways. 

The final results obtained accordingly were fit for visualization, as shown in Figs. \ref{fig:ResultsWallcamera}(b) and 2(c). 

In DIF \cite{dif}, all operations were performed in the native linear space of values, as opposed to what was done in the Wallcamera. The operations in DIF specific for video data were as follows. First, a reference frame was created by averaging all frames from a temporal sub-interval of the video during which no person or object of interest was present in the scene. This differs from the full-video temporal average that is generically used in the Wallcamera for subtraction. Next, difference images were computed as the pixel-wise differences between every video frame and the reference frame. Averaging of the resulting temporal images was then performed through both spatial and temporal filtering. Specifically, the difference video frames were spatially filtered with a 2D Gaussian filter, and temporally filtered with a uniform temporal window. Following these filtering operations, the contrast was normalized similarly as in the Wallcamera, except that only the positive or negative values of the resulting difference frames were retained before normalization.  
This was done to maximize contrast and achieve optimal visualization of the extracted signals for DIF,
as demonstrated by the results in Figure \ref{fig:ResultsDIF}(b).

\subsection{Parameter estimation}
\label{subsect: ParameterEstimation}

Parameter estimation is a fundamental concept in inverse problems. It involves deriving a physical model from a finite set of observations or measurements, which may contain errors, and determining quantitative or categorical characteristics (i.e., parameters) that describe the objects of interest in the model, based on the available data.

In terms of parameter estimation, the Wallcamera and DIF bear some similarities in that 1) both achieved the retrieval of non-trivial scene-related parameters via difference information, and 2) both relied on the same type of differential information in nature. However, they used the differential information in distinct ways, and they involved the estimation of distinct types of parameters.

The parameters of interest estimated in DIF were of the quantitative type: essentially they were an intruder's approximate locations over time. In effect, these location estimates depicted his movement, his body information such as height and width, and the color of his clothing.
These parameters were directly inferred visually from the difference images that DIF extracted. 
The joint spatio-temporal filtering was a purpose-built design choice for this setting, delivering high-quality results.

In the Wallcamera, the  parameters of interest were mostly categorical: parameter estimation mostly focused on recognising different activity categories, 
except for the determination of the number of people in a room.

The Wallcamera estimated the scene parameters via a post-processed version of the extracted difference images, as opposed to directly exploiting them as such.
Specifically, the set of difference images (which were 3D data defined in two spatial dimensions plus time) was projected into two 2D maps that summarize the essential spatial dimensions of the motion in a separate manner: a horizontal space-time plot and a vertical space-time plot.

While such space-time plots lack the direct visual interpretability that DIF benefits from in terms of resolved spatial information, they offer some advantages. First, they summarize spatio-temporal variations into a single 2D plot, conveniently visualizing the presence of movement. Second, compared to a series of video frames, the dimensionally reduced information in these 2D plots provides a compact and computationally efficient input for a learning model.

\subsection{Machine learning}

The methods used to estimate the parameters of interest from the extracted difference-image data
also differ. In the Wallcamera, two separate convolutional neural networks (CNNs) were built for deducing the number of people, and for activity recognition, respectively. These networks took the post-processed difference images, summarized into two 2D space-time plots 
as described above, as input. These plots contained the amplified tale-telling signals in a dimensionally reduced form.

In Bourquard and Yan \cite{dif}, no neural networks were implemented to perform activity recognition. However, they demonstrated that simple signal processing could achieve similarly compelling results. 

Finally, the difference images obtained as in DIF may also be used as input to CNNs, with the potential to produce activity recognition results similar to those in the Wallcamera. Bourquard and Yan \cite{dif} did envision the use of neural networks for boosting differential signal extraction and other capabilities, as quoted in the following.

\begin{enumerate}

\item \textit{``more advanced algorithms could involve better noise-reduction techniques and the use of machine learning, such as artificial neural networks, to identify specific evidence stemming from particular objects in a scene"}

\item \textit{``Enhanced analysis capabilities would in turn allow to explore the applicability of DIF to more complex scenarios and perhaps recover visual evidence in cases where the objects of interest have less impact on the scene images"}

\end{enumerate}

%% file: sections/conclusion.tex
\section{Conclusions}\label{conc}

Sharma et al \cite{iccv21} presented some interesting and innovative research. However, their key insight was the same as that in our paper \cite{dif}
published several years earlier, which reported the invention of \textit{differential imaging forensics}.
Furthermore, there are other significant conceptual and methodological similarities between the two works.
For example, both used 
a similar signal-extraction approach to capture temporal differences between video frames, 
and a subsequent signal amplification  
that applied spatial filtering to increase the 
signal-to-noise ratio. Moreover, the experimental setups were nearly identical,
both following a NLOS setting where the camera was pointed towards a wall, and where the visual cues of interest to be recovered were invisible to a human observer. 

DIF was the first to propose and successfully demonstrate activity recognition in such an experimental setup.
The actual innovation of the Wallcamera  
was activity recognition at a finer granularity than what DIF demonstrated. 
The Wallcamera used convolutional neural networks. 
However, DIF only used simple signal processing, 
without resorting to the power of deep learning.

Unfortunately,  
the Wallcamera \cite{iccv21} did \emph{not} cite the prior art of 
Bourquard and Yan \cite{dif}, which was circulated
via \url{arxiv.org} and  
shared with the same community.
The research on DIF was also presented in invited talks at some high-profile institutions across the world, including Cambridge, TU Damsdat and Aarhus, to name a few. 
On a positive note, the Wallcamera in a sense provided an independent and valuable validation of the ideas and insights behind DIF, although from a different perspective.

Deeply rooted in security and forensics, 
DIF entails more than a capability for activity recognition. As detailed in our paper series documenting its invention \cite{dif,yan2017poster,dif2022}, DIF offers a range of additional forensic applications, e.g.  
the recovery of a photographer's personal identifiable information from a single photo,  
the detection of image tampering, and the mitigation of deepfakes. Though compelling, activity recognition merely constitutes a single use case which we initially  
envisaged and demonstrated. DIF deserves due credit and attention in computer vision, security, forensics and beyond.

\section*{Postscripts} 

During the finalization of this manuscript, we came across Medin et al \cite{freeman2}, entitled `\textit{Can Shadows Reveal Biometric Information}?', a newer  
paper from the Wallcamera team. 
We note some interesting parallels and differences between DIF \cite{dif} and Medin et al \cite{freeman2} as follows. 

The primary aim of DIF was to uncover both biometrics (such as body height and width) and non-biometric characteristics (such as clothing color) based on the subtle and invisible visual impacts created in the scene by human subjects outside of the FoV. In a similar vein, a main contribution of Medin et al \cite{freeman2} was, as quoted, 
``\textit{a timely biometric leakage question, which [they] formulate as a novel NLOS imaging problem of extracting an individual’s identity from subtle, indirect shadow phenomena}". They defined biometric information as ``\textit{any information that might be used to reveal an individual’s identity, in
whole or in part.}"

However, DIF is sufficiently generic to work with visual impacts beyond just 
shadows. These impacts may originate from various types of light interactions involving the scene and the object of interest, including absorption (as with shadows), reflection, or other forms of light interaction, as long as enough information is captured by the difference images that DIF recovers.
Accordingly, in the context of DIF, we define such a generic set of visual impacts as an \emph{impact profile}, which encompasses visual impacts created by any type of light interaction or setup. The DIF video experiment constituted an explicit NLOS setting where information from a human subject could be extracted successfully from the impact profile. 

There are two additional differences in 
origin and characteristics between the impact profile considered in DIF \cite{dif} and the shadow in Medin et al. \cite{freeman2}. First, the impact profile in DIF was cast directly by the human subject. By contrast, Medin et al. \cite{freeman2} introduced an additional layer of indirection. Specifically, their setup
involved an added occluder in the scene, which sit between the wall and the human subject that was outside of the FoV. The shadow of the occluder was cast on the wall, created by the light reflecting from the human face. The shadow considered in their setup was thus not cast directly by the human subject, but by the occluder. 
Since the light creating the shadow was reflected from the human subject, the shadow's properties were subtly modulated by the human face. Medin et al. aimed to recover facial information from this modulated shadow.

Second, in DIF studies, the impact profile was usually invisible to the human eye, whereas the shadow examined by Medin et al. \cite{freeman2} was clearly visible. However, most of the biometric information detected by Medin et al. \cite{freeman2} was only present in the penumbra regions of the shadow, while the umbra regions carried little such information.

Overall, there are interesting and clear parallels between DIF and the work of Medin et al \cite{freeman2} -- they bear some similarities, and they also differ substantially.  
Like Sharma et al \cite{iccv21}, Medin et al \cite{freeman2} did not cite Bourquard and Yan \cite{dif}.